\documentclass[11pt]{article}
\usepackage[margin=1in]{geometry}
\usepackage{amsmath,amssymb,amsthm,bm,mathtools}
\usepackage{booktabs}
\usepackage{graphicx}
\usepackage{float}
\usepackage[colorlinks=true,linkcolor=blue,citecolor=blue,urlcolor=blue]{hyperref}

\newtheorem{proposition}{Proposition}

\title{Duplicate-Aware Shift-and-Lift Carleman Linearization:\\Structure, Complexity, and Comparative Evaluation}
\author{
Takaki Akiba\\
The University of Tokyo
\and
Youhi Morii\\
Institute of Fluid Science, Tohoku University
}
\date{\today}

\begin{document}

\maketitle

\begin{abstract}
The primary objective of this study is to remove duplicated monomial contributions that proliferate in Carleman linearization as state dimension and truncation order increase. To do so, we adopt a shift-and-lift architecture, since it exposes repeated exponent targets and allows duplicate-aware coefficient coalescing during lifted-operator assembly. This architecture also makes high-order truncation practical, but that regime intensifies local convergence and closure sensitivity for higher-order nonlinearities. We therefore pair shift-and-lift with a moving-center expansion so that shift and lift are updated jointly around evolving local centers, improving validity of the truncated model along the trajectory. The resulting workflow combines symmetry-reduced monomial bases, packed exponent-key indexing, and sparse triplet coalescing to preserve truncated affine dynamics while reducing index-resolution overhead and write-path irregularity. We analyze variable growth, preprocessing complexity, and truncation-induced error mechanisms, and we compare against Jacobian linearization through fixed-step error, admissible step size, and cost-at-target-accuracy criteria. Two benchmarks (bilinear driver and logistic interaction) show convergence under refinement for both approaches, with regime-dependent accuracy gains for the proposed method rather than universal superiority.
\end{abstract}

\section{Introduction}
Carleman linearization converts polynomial nonlinear dynamics into an infinite linear system on lifted monomials \cite{carleman1932}. Because finite truncations grow rapidly with dimension and order and are typically reliable only over local validity regions, Carleman models are often too costly for routine large-scale numerical integration; instead, they have been used more frequently in control and estimation settings for mildly nonlinear dynamics around operating points \cite{weber2016,rauh2009,rotondo2022}. Its relevance has recently expanded in quantum-oriented workflows, where core algorithmic primitives are naturally linear-algebraic and often built around linear-system solves \cite{hhl2009}. Carleman embeddings therefore provide a direct bridge from nonlinear dynamics to quantum-compatible operator forms.

Recent work shows a clear trend from proof-of-concept usage toward regime-explicit algorithm design. A key milestone is the dissipative-regime quantum algorithm of Liu \emph{et al.}, which established explicit truncation and complexity guarantees for quadratic nonlinear ODEs under strong stability assumptions \cite{liu2021}. Later analyses improved accuracy dependence and reduced overhead in linear-ODE subroutines \cite{krovi2023,costa2025}. In parallel, the admissible dynamical regime has expanded beyond strict dissipativity: resonance-based theory now provides convergence guarantees in broader non-dissipative settings \cite{wu2025}. This shift from ``dissipative only'' to ``structure-conditioned'' guarantees is central to the current quantum-Carleman literature.

A second trend is deeper convergence theory for the classical embedding itself. Beyond local finite-section analyses, recent studies propose piecewise/global chart-switching Carleman embeddings to address trajectories that leave a single convergence neighborhood \cite{novikau2025}. Related functional-analytic work extends well-posedness and convergence results to parabolic PDE settings in infinite-dimensional Hilbert spaces, while separating discretization and linearization error channels \cite{heinzelreiter2025}. Together, these results sharpen when and why truncated linear embeddings can remain reliable.

A third trend is methodological hybridization. Application-oriented studies now test Carleman routes in domain models relevant to quantum simulation pipelines, including chemical kinetics and fluid dynamics \cite{akiba2023,sanavio2024}. At the same time, basis-augmented and comparator frameworks---such as Carleman-Fourier constructions for periodic vector fields and Koopman-spectral alternatives---are being explored to improve behavior outside strictly polynomial regimes \cite{motee2026,shi2024}. Perspective-level assessments consequently frame nonlinear quantum simulation as a co-design problem between approximation structure, truncation control, and hardware constraints \cite{tennie2025}.

A recurring conclusion across these strands is that asymptotic quantum advantages can be masked by classical front-end costs if lifted operators are expensive to assemble or repeatedly rebuilt.

The present work addresses precisely this front-end bottleneck for moving-center shift-and-lift construction. In end-to-end pipelines, the final sparse linear solve is itself a major challenge and a natural candidate for quantum acceleration; however, preconditioning and related conditioning/assembly tasks are not generally handled efficiently by current quantum workflows and must still be executed classically. Consequently, practical performance requires eliminating classical bottlenecks one by one. The specific bottleneck targeted here is repeated generation and late reconciliation of duplicate exponent targets. We introduce a duplicate-aware assembly workflow based on symmetry-reduced monomial bases, packed exponent-key indexing, and sparse triplet coalescing during construction rather than post hoc correction.

The manuscript adopts a deliberately balanced perspective. The goal is not to claim universal superiority over first-order Jacobian linearization, but to identify regimes where added representational richness outweighs truncation-closure and conditioning penalties. This framing motivates both the methodological details and the comparative experiments reported below.

\section{Problem Setting and Notation}
Consider
\begin{align}
\dot{x}=f(x), \qquad x\in\mathbb{R}^n,
\end{align}
with polynomial right-hand side represented in a monomial basis $\mathcal{Y}$ up to degree $P$:
\begin{align}
\dot{x}=A_{XY}y(x).
\end{align}
Two basis sets are used:
\begin{align}
\mathcal{Y}&=\{x^\beta:1\le |\beta|\le P\}, \\
\mathcal{Z}&=\{x^\alpha:1\le |\alpha|\le Q\},
\end{align}
with cardinalities $n_Y=|\mathcal{Y}|$ and $n_Z=|\mathcal{Z}|$. Here, $x^\beta=\prod_{j=1}^n x_j^{\beta_j}$ and $|\beta|=\sum_j \beta_j$.

Both bases are symmetry-reduced monomial sets generated by combinations-with-repetition, and degree-$0$ is excluded in both $\mathcal{Y}$ and $\mathcal{Z}$. The input polynomial operator is therefore
\begin{align}
A_{XY}\in\mathbb{R}^{n\times n_Y},
\end{align}
with one row per state equation and one column per nonconstant monomial in $\mathcal{Y}$. Throughout, we assume $P\le Q$ and use precomputed basis-index maps for repeated operator assembly.

\section{Duplicate-Aware Shift-and-Lift Construction}

As $(n,P,Q)$ increase, multiple combinatorial term-generation paths map to identical monomial exponents, so duplicate contributions become a primary assembly bottleneck. The shift-and-lift architecture is introduced to expose this structure explicitly and enable duplicate-aware coalescing in the lifted dynamics. Because this same architecture can support higher truncation orders, convergence and closure issues for higher-order nonlinear terms become more pronounced in fixed-center expansions. To mitigate this effect, the construction is carried out in a moving-center manner, so shift and lift are performed jointly around updated local centers.

\subsection{Shift Stage}
For shift point $x_0$, expansion of $(x+x_0)^\alpha$ yields
\begin{align}
(x+x_0)^\alpha
=\sum_{0\le r\le \alpha}
\left(\prod_{j=1}^n\binom{\alpha_j}{r_j}\right)x^r x_0^{\alpha-r}.
\end{align}
The shift stage uses this identity in direct form (without explicitly materializing a dense shift matrix). Terms with $|r|=0$ contribute to $b_Y$, while terms with $|r|\ge 1$ contribute to $A_{\mathrm{shift}}$. Hence
\begin{align}
b_Y(i)&=\sum_{\alpha\in\mathcal{Y}}A_{XY}[i,\alpha]x_0^\alpha,\\
A_{\mathrm{shift}}[i,r]
&=\sum_{\alpha\in\mathcal{Y}}A_{XY}[i,\alpha]
\left(\prod_{j=1}^n\binom{\alpha_j}{r_j}\right)x_0^{\alpha-r},
\end{align}
for $r\in\mathcal{Y}$, and
\begin{align}
\dot{x}_i=b_Y(i)+\sum_{r\in\mathcal{Y}}A_{\mathrm{shift}}[i,r]y_r.
\end{align}

\subsection{Lift Stage}
For $z_\alpha=x^\alpha$ with $\alpha\in\mathcal{Z}$,
\begin{align}
\dot{z}_\alpha
=\sum_{i=1}^n\alpha_i x^{\alpha-e_i}\dot{x}_i.
\end{align}
Substitution gives
\begin{align}
\dot{z}_\alpha
&=\sum_{i=1}^n\sum_{\beta\in\mathcal{Y}}\alpha_iA_{\mathrm{shift}}[i,\beta]x^{\alpha-e_i+\beta}
+\sum_{i=1}^n\alpha_i b_Y(i)x^{\alpha-e_i},
\end{align}
followed by truncation to degrees $1\!:\!Q$. A contribution is kept only if its target exponent is in $\mathcal{Z}$; otherwise it is dropped (or, in an affine-closure variant, folded into $b_Z$ through evaluation at $x_0$). The resulting form is
\begin{align}
\dot{z}=A_{ZZ}z+b_Z.
\end{align}

Because degree-$0$ is excluded from $\mathcal{Z}$, the $b_Y$-driven term is handled in two cases: if $|\alpha-e_i|=0$, it contributes directly to $b_Z$; if $1\le |\alpha-e_i|\le Q$, it is inserted into $A_{ZZ}$ at the corresponding monomial index.

\subsection{Duplicate-Aware Assembly Principle}
Distinct tuples $(\alpha,i,\beta)$ may map to the same exponent target $\gamma=\alpha-e_i+\beta$. The proposed assembly resolves target indices by packed exponent keys,
\begin{align}
\kappa(\gamma)=\sum_{j=1}^n \gamma_j 2^{(j-1)b},
\qquad
b=\left\lceil\log_2(2Q+1)\right\rceil,
\end{align}
with a key-to-column lookup map in $\mathcal{Z}$ (subject to the finite-word packing constraint $nb\le 64$). Duplicate triplets are then coalesced during sparse compression. This preserves exact truncated coefficients up to floating-point summation order while avoiding repeated combinatorial index search.

\begin{proposition}[Exactness under fixed truncation]
For fixed $(P,Q)$ and fixed shift $x_0$, key-resolved triplet assembly yields the same truncated affine pair $(A_{ZZ},b_Z)$ as term-by-term expansion in exact arithmetic, differing only by floating-point summation order.
\end{proposition}

\section{Complexity and Variable Growth}

\subsection{Counting Conventions and Variable Growth}
Two counting conventions are standard. Under tensor-product ordering,
\begin{align}
N^{\mathrm{tensor}}(Q)=\sum_{k=1}^Q n^k=\frac{n^{Q+1}-n}{n-1}, \qquad n>1.
\end{align}
Under symmetry-reduced monomials,
\begin{align}
N^{\mathrm{sym}}(Q)=\sum_{k=1}^Q\binom{n+k-1}{k}=\binom{n+Q}{Q}-1.
\end{align}
In explicit two-layer shift-and-lift form, both $\mathcal{Y}$ and $\mathcal{Z}$ are represented, so state dimension exceeds that of a single-layer standard Carleman representation under either convention.

\subsection{Computational Complexity Decomposition}
Let $C_\phi$ denote average cost to evaluate one basis component from $x$. Lifted-vector preparation scales as $\mathcal{O}(N_{\mathrm{std}}C_\phi)$ for single-layer Carleman and as $\mathcal{O}((n_Y+n_Z)C_\phi)$ for explicit two-layer shift-and-lift.

For matrix preparation, define generated-term counts $T_{\mathrm{std}}$, $T_{\mathrm{shift}}$, $T_{\mathrm{lift}}$ and merged nonzero counts $U_{\mathrm{std}}$, $U_{\mathrm{ours}}$. A practical model is
\begin{align}
\text{standard prep} &\sim \mathcal{O}(T_{\mathrm{std}})\ \text{updates} + \mathcal{O}(U_{\mathrm{std}})\ \text{writes},\\
\text{proposed prep} &\sim \mathcal{O}(T_{\mathrm{shift}}+T_{\mathrm{lift}})\ \text{updates} + \mathcal{O}(U_{\mathrm{ours}})\ \text{writes}.
\end{align}
The key reduction appears in write-path irregularity: duplicate-aware accumulation ties final writes to unique targets instead of raw generated terms.

Memory behavior follows the same pattern. Naive triplet-heavy workflows may exhibit peak temporary storage scaling with generated contributions; early key-based merging keeps temporary state closer to unique structure size (plus cache metadata), improving locality and reducing peak RAM pressure.

When the same combinatorial structures are reused across multiple $(A_{XY},x_0)$ instances, one-time cache construction is amortized, further improving effective throughput.

\subsection{Closure Residual and Local Error Mechanism}
With finite-order closure,
\begin{align}
\dot{y}=Ay+b+r(y),
\end{align}
where $r(y)$ aggregates neglected terms. Under implicit Euler,
\begin{align}
(I-\Delta t A)y_{\mathrm{full}}^{1} &= y^0 + \Delta t b + \Delta t r(y_{\mathrm{full}}^{1}),\\
(I-\Delta t A)y_{\mathrm{lin}}^{1} &= y^0 + \Delta t b,
\end{align}
so
\begin{align}
\delta y^1=(I-\Delta t A)^{-1}\Delta t\,r(y_{\mathrm{full}}^{1}).
\end{align}
Therefore error growth is jointly controlled by residual magnitude and resolvent amplification. Small-step behavior near the shift origin can remain benign, whereas larger steps may amplify closure effects through both residual growth and conditioning deterioration.

\section{Comparison with Jacobian Linearization}

\subsection{Conceptual Contrast}
Jacobian linearization applies the first-order local approximation
\begin{align}
f(x)\approx f(x^*) + J(x^*)(x-x^*).
\end{align}
Its strength is low per-step complexity and broad local robustness. Shift-and-lift retains higher-order structure through augmented coordinates, potentially improving fidelity when dominant nonlinearities are well represented by retained monomials.

The trade-off is explicit: higher representational richness versus increased dimensionality, closure error, and conditioning sensitivity. Consequently, superiority is expected to be problem-dependent rather than universal.

\subsection{Quantitative Comparison Criteria}
Let
\begin{align}
E(\Delta t)=\max_{0\le n\le N}\|x_n^{\mathrm{approx}}-x_n^{\mathrm{ref}}\|,
\qquad
\Delta t_{\max}(e)=\max\{\Delta t\mid E(\Delta t)\le e\}.
\end{align}
Define the step-size gain factor
\begin{align}
R(e)=\frac{\Delta t_{\max}^{\mathrm{lift}}(e)}{\Delta t_{\max}^{\mathrm{Jac}}(e)}.
\end{align}
In addition, fixed-step error curves, fixed-error computational cost, and conditioning or boundedness diagnostics should be reported jointly.

\subsection{Balanced Positioning Statement}
The appropriate manuscript-level claim is that shift-and-lift can outperform Jacobian linearization when dominant nonlinearities align with a moderate-order lifted basis and lifted dynamics remain bounded and well conditioned. When truncation closure or conditioning penalties dominate, Jacobian linearization can be comparable or superior.

Accordingly, the central methodological question is not whether higher-order lifting is intrinsically superior, but whether representational gains exceed closure and conditioning penalties at the target tolerance. Performance claims should therefore be supported jointly by $R(e)$, fixed-step error curves, conditioning or boundedness diagnostics, and cost-at-target-error comparisons.

\section{Demonstrations}

\subsection{Demo 1: Bilinear Driver Benchmark}
The first benchmark is carried out with a bilinear driver system described by
\begin{align}
\dot{x}=-\lambda_x x + kuv,
\qquad
\dot{u}=-\lambda_u u,
\qquad
\dot{v}=-\lambda_v v.
\end{align}
The numerical study first generates a high-accuracy reference trajectory with an adaptive integrator, then compares two implicit-Euler-type approximations on a matched grid: a moving-center shift-and-lift solver, and a Jacobian baseline using per-step linear solves of $(I-\Delta t J_n)\Delta x_n=\Delta t f(x_n)$.

The lifted solver rebuilds $(A_{ZZ},b_Z)$ around the current state and solves
\begin{align}
(I-\Delta t A_{ZZ})z^{n+1}=\Delta t b_Z,
\end{align}
with state reconstruction from lifted increments. For each run, the study records trajectory outputs, final operators, and error tensors relative to the sampled reference, together with Frobenius-norm error summaries and comparison plots for $(x,u,v)$.

Parameter sweeps currently cover $\deg Z\in\{3,4,5\}$ and $n_{\mathrm{eval}}\in\{101,201\}$. These outputs provide the empirical basis for evaluating fixed-step accuracy, step-size sensitivity, and practical trade-offs between lifted and Jacobian approaches.

\begin{figure}[H]
  \centering
  \includegraphics[width=0.75\linewidth]{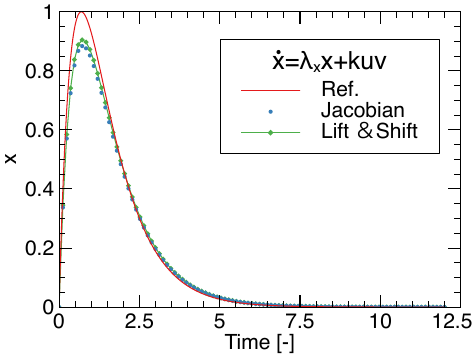}
  \caption{Trajectory comparison for Demo 1 (state $x$): reference solution versus Jacobian and shift-and-lift approximations.}
  \label{fig:demo1_x_traj}
\end{figure}

Figure~\ref{fig:demo1_x_traj} presents the time history of the state $x$ for the reference, Jacobian, and shift-and-lift models. Both reduced models capture the dominant transient profile and long-time decay behavior. The main discrepancy is concentrated near the early peak, where the reduced models slightly under-estimate amplitude relative to the reference trajectory.

\begin{figure}[H]
  \centering
  \includegraphics[width=0.75\linewidth]{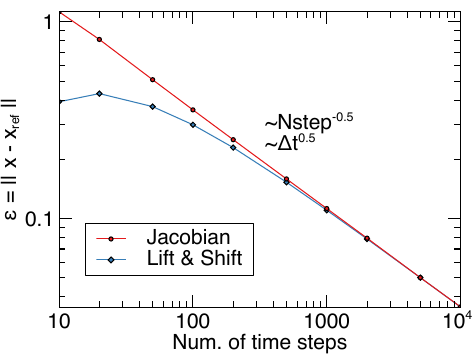}
  \caption{Error scaling in Demo 1 versus number of time steps. Both methods converge with approximately first-order behavior; shift-and-lift shows a small advantage in this case.}
  \label{fig:demo1_err_steps}
\end{figure}

Figure~\ref{fig:demo1_err_steps} shows error magnitude as a function of the number of time steps. Both methods exhibit consistent error reduction under temporal refinement, and the observed slope is approximately first-order with respect to $\Delta t$, as expected for implicit-Euler-type updates. Over most of the sampled range, the shift-and-lift curve is slightly below the Jacobian curve, indicating a modest but systematic accuracy gain in this benchmark.

\subsection{Demo 2: Logistic Interaction Benchmark}
The second benchmark uses a coupled logistic-interaction model,
\begin{align}
\dot{x} &= a x - b x^2 - cxy,\\
\dot{y} &= d y - e y^2 - fxy,
\end{align}
with $(a,b,c,d,e,f)=(1.0,1.0,0.5,0.8,1.2,0.4)$ and initial condition $(x(0),y(0))=(0.2,0.3)$ over $t\in[0,20]$. This system provides a compact test with both self-limiting quadratic terms and cross-coupling, so it is useful for checking how well the lifted representation captures nonlinear interaction dynamics in two dimensions.

The numerical workflow follows the same solver structure as Demo~1 and therefore omits repeated derivation details. In short, the study constructs a polynomial operator $A_{XY}$ for degree-$2$ monomials, then compares: (i) a moving-center shift-and-lift solve with truncated degree $\deg Z\in\{2,3,4\}$, and (ii) a Jacobian baseline on the same time grid. A high-accuracy reference trajectory is generated with an adaptive high-order ODE solver at tight tolerances, and all reduced-model errors are evaluated against the reference sampled on matching nodes.

Parameter sweeps are performed over $n_{\mathrm{eval}}\in\{11,21,51,101,201,501,1001,2001,5001,10001\}$. For each run, trajectories, final operators, and norm summaries are recorded. This supports direct comparison of state accuracy, convergence trends with temporal refinement, and the sensitivity of lifted performance to truncation order.

\begin{figure}[H]
  \centering
  \includegraphics[width=0.75\linewidth]{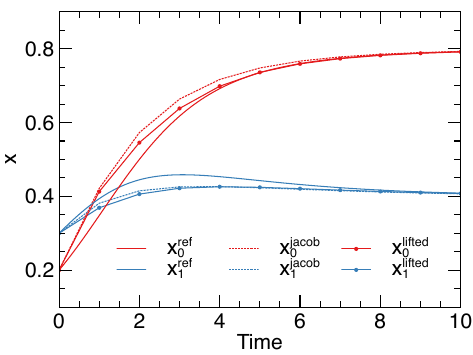}
  \caption{Trajectory comparison for Demo 2 (logistic interaction): reference, Jacobian, and shift-and-lift solutions for both state components.}
  \label{fig:demo2_xy_traj}
\end{figure}

Figure~\ref{fig:demo2_xy_traj} shows representative trajectories for both states. The lifted and Jacobian solutions track the reference dynamics well across the interval, with visible differences concentrated during the earlier transient where nonlinear coupling is strongest.

\begin{figure}[H]
  \centering
  \includegraphics[width=0.75\linewidth]{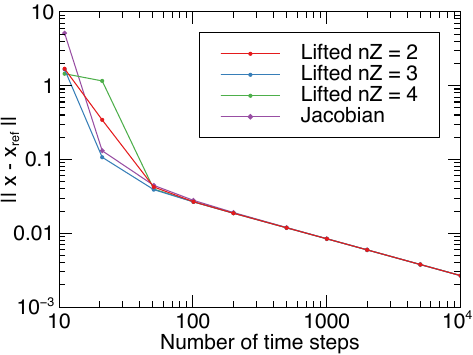}
  \caption{Error scaling in Demo 2 versus number of time steps for Jacobian and lifted models with $n_Z\in\{2,3,4\}$.}
  \label{fig:demo2_err_steps}
\end{figure}

Figure~\ref{fig:demo2_err_steps} reports error magnitude versus number of time steps on log--log axes. As in Demo~1, all curves decrease with refinement, and the lifted method with higher $\deg Z$ shows improved coarse-grid behavior before curves merge in the fine-grid regime.

\section{Conclusion}
This study presents a construction-grounded shift-and-lift Carleman framework in which duplicate combinatorial contributions are handled through key-resolved sparse assembly. The resulting pipeline preserves the intended truncated affine model while improving practical assembly regularity through cached basis structures, packed-key lookup, and compressed sparse output formation.

The analysis and demonstrations support a balanced conclusion. Shift-and-lift provides clear structural advantages for representing nonlinear interactions and can deliver modest accuracy benefits at comparable discretizations in the tested bilinear and logistic benchmarks. At the same time, performance remains problem-dependent: truncation closure, lifted-system conditioning, and operating step size govern whether these advantages outweigh the added lifted dimension and preprocessing burden. For that reason, future comparative studies should continue to report accuracy, stability, and computational cost under common tolerances and grids.

\section*{Supplementary Material}
The implementation code set associated with this manuscript is available at:
\url{https://github.com/takakiba/duplicate-aware-carleman-linearization.git}.

\end{document}